\documentclass[preprint,pra,showpacs,nofootinbib]{revtex4}
\usepackage{appendix}
\usepackage{float,epsfig}
\usepackage{dcolumn}
\usepackage{bm}
\usepackage{graphicx}
\usepackage{dcolumn}
\usepackage{bm}
\usepackage{amsmath,amssymb,amsthm}
\usepackage[colorlinks=true,linkcolor=blue]{hyperref}
\begin{document}
\title{\bf The far-zone interatomic Casimir-Polder  potential between two ground-state atoms outside a Schwarzschild black hole }
\author{ Jialin Zhang $^{1}$  and Hongwei Yu $^{2,1}$
} \affiliation{$^1$ Institute of Physics and Key Laboratory of Low
Dimensional Quantum Structures and Quantum
Control of Ministry of Education,\\
Hunan Normal University, Changsha, Hunan 410081, China \\
$^2$ Center for Nonlinear Science and Department of Physics, Ningbo
University,  Ningbo, Zhejiang 315211, China}

\date{\today}

\begin{abstract}
   Based on the idea that the  vacuum fluctuations of  electromagnetic fields
 can induce instantaneous  correlated  dipoles, we study the far-zone Casimir-Polder potential between two atoms  in the
 Boulware, Unruh and Hartle-Hawking vacua
 outside a Schwarzschild black hole.  We show that, at
spatial infinity, the Casimir-Polder potential  in the Boulware
vacuum is similar to that in the Minkowski vacuum in flat spacetime
with  a behavior of $R^{-7}$, so is  in the Unruh vacuum as a result
of the backscattering of the Hawking radiation from the black hole
off the spacetime curvature.  However, the interatomic
Casimir-Polder potential in the Hartle-Hawking vacuum behaves like
that in a  thermal bath at the Hawking temperature. In the region
near the event horizon of the black hole, the modifications caused
by the space-time curvature make the interatomic Casimir-Polder
potential smaller in all three vacuum states.
\end{abstract}
 \maketitle
\section{Introduction}
The Casimir effect which can be considered as one of the
macroscopical observable phenomena originating  from the vacuum
field fluctuations was firstly discussed by Casimir  in 1948
~\cite{Casimir-1}. Casimir predicted that vacuum fluctuations  give
rise to an attractive force between two neutral conducting plates at
rest. In the same year, Casimir and Polder also began the pioneering
work on the retarded dispersion interaction between two atoms(or
molecules)~\cite{Casimir-Polder}. For atoms having a dominant
transition with frequency $\omega_0$ between the ground and first
excited states, they showed that the interaction between the two
atoms reduced to the London limit of the van der Waals interaction
in the near zone, i.e., $R^{-6}$ dependence for small separations
($R\omega_{0}\ll1$). In contrast, the interaction energy decays like
$R^{-7}$ in the far zone~\cite{Casimir-Polder}. So far, the Casimir
and Casimir-Polder forces have been measured  with remarkable
precision in experiments~\cite{Sukenik-ex}.

Since  space-time geometry and the presence of boundaries can affect
vacuum field fluctuations, it is  expected that  the Casimir-Polder
interaction will be modified in these circumstances. In this regard,
the Casimir-Polder interaction between two atoms placed near the
conducting plate was studied by Spagnolo et al ~\cite{Spagnolo}. A
natural question along that line is what happens when the two-atoms
system is placed in curved spacetime rather than a flat spacetime. This is
what we are going to do in the present paper, i.e.,  we are going to
investigate the Casimir-Polder potential between two neutral but
polarizable atoms outside a spherically symmetric black hole.
  Let us note, as examples of related effects that also arise as a result of the modification of
  vacuum fluctuations due to the presence of spacetime curvature,
that the Lamb shift of a static atom~\cite{zhy-1,zhy-2} and the
Casimir-Polder-like force on it~\cite{zhjl} outside a Schwarzschild
black hole have recently been studied.

There are numerous methods aimed at obtaining the Casimir-Polder
potential, such as those using two-transverse-photon exchange within
perturbation theory~\cite{margenau-book,Craig-book}, consideration
of the changes in zero-point energy~\cite{Boyer} , radiative
reaction ~\cite{PMilonni}, evaluation of energy shifts in the
Heisenberg picture~\cite{Power}, the method based on spatial vacuum
field correlations\cite{Power-Thirunamachandran,Spagnolo},  the
response theory ~\cite{Barash-book}  and so on. A general treatment
within a relativistic framework is reviewed by Feinberg and
Sucher~\cite{Feinberg}. Our calculation of the interatomic
Casimir-Polder potential is based upon the method of equal-time
spatial vacuum field correlations which  can simplify some
calculations in some complex external environment. The main idea
based on the vacuum spatial correlations can be narrated as that the
vacuum fluctuations of the electromagnetic field  induce
instantaneous correlated dipole moments on the two atoms and
the Casimir-Polder potential energy  can be  obtained  by
calculating the classical interaction between the two correlated
induced dipoles~\cite{Power-Thirunamachandran,Spagnolo}.

The paper is organized as follows. In the next section, we will give
the basic formula of interatomic Casimir-Polder potential between
the two ground-state atoms in the far zone. Then we will calculate
Casimir-Polder potential caused by induced instantaneous atomic
dipoles generated by electromagnetic field fluctuations in the
Boulware vacuum~\cite{Boulware}, Unruh vacuum ~\cite{Unruh}and
Hartle-Hawking vacuum ~\cite{Hartlehawking}, In Secs. III, IV, and V
respectively. Finally, we will conclude in Sec. VI.

\section{ The field spatial correlation function and the interatomic Casimir-Polder  potential}
Within the dipole approximation, the Hamiltonian of a system
composed of two atoms A and B interacting with external radiation
fields in the multipolar scheme  can be written as
\begin{equation}
H=H_F+H^A_{atom}+H^B_{atom}-{\bf{\mu}}_A\cdot{\bf{D}}({\bf
{r}}_A)-{\bf{\mu}}_B\cdot{\bf{D}}({\bf {r}}_B)\;,
\end{equation}
where ${\bf{D}}({\bf {r}}_A)=\sum{\bf{D}}(\omega_{\bf{ k}},{\bf
{r}}_A)$ denotes the transverse displacement electric field operator
at the point ${\bf {r}}_A$ and ${\bf{\mu}}_A$ (or ${\bf{\mu}}_B$)
indicates the electric dipole operator of atom A ( or B). For the
two atoms which are fixed at the certain locations in a space-time,
the vacuum fluctuations of the electromagnetic field induce
instantaneous correlated  dipole moments on them as a result of the
spatially correlated vacuum fluctuations. The Casimir-Polder
potential energy then  can be considered as the (classical)
interaction between the two correlated induced dipoles.
 The induced dipole moments caused by the
vacuum fluctuations usually can be written  as~\cite{Power-Thirunamachandran,Spagnolo}:
$
\mu_l(\omega_{\bf {k}})=\alpha(\omega_{\bf {k}})D_l(\omega_{\bf
{k}},{\bf{r}})\;,
$
where
\begin{equation}\alpha(\omega_{\bf
{k}})=\frac{2}{3}\sum_m\frac{E_{m0}\mu_{m0}^2}{E_{m0}^2-\omega_{\bf
{k}}^2}\end{equation} is the atomic dynamical isotropic
polarizability (here $E_{m0}=E_m-E_0$ and $\mu_{m0}$ denote the
matrix elements of the atomic dipole moment operator). For the
case of a two-level atom,  the isotropic polarizability can be
written as
\begin{equation}
\alpha(\omega_{\bf
{k}})=\frac{2\omega_0\mu^2}{3(\omega_0^2-\omega_{\bf
{k}}^2)}\;.\end{equation}

    Therefore, the interatomic Casimir-Polder  potential of two ground-state atoms  reads
 ~\cite{Power-Thirunamachandran,Spagnolo}
\begin{equation}\label{VAB}
V_{AB}=\int\sum_{ij}\alpha_A(\omega_{\bf {k}})\alpha_B(\omega_{\bf
{k}})\langle {D_i(\omega_{\bf{ k}},{\bf{r_A}})}{D_j(\omega_{\bf
{k}},{\bf {r_B}})}\rangle{V}_{ij}(\omega_{\bf{ k}},R)d\omega_{\bf
{k}}\;,
\end{equation}
where $\langle {D_i(\omega_{\bf{ k}},{\bf{r_A}})}{D_j(\omega_{\bf
{k}},{\bf {r_B}})}\rangle $ is the equal-time spatial correlation
function of the electric field in the vacuum state and
${V}_{ij}(\omega_{\bf {k}},R)$ is the classical electrostatic interaction energy between two dipoles oscillating at frequency
$\omega_{\bf{ k}}$~\cite{McLone}
\begin{equation}\label{Vij}
{V}_{ij}(\omega_{\bf
{k}},R)=(\delta_{ij}-3\hat{R_i}\hat{R_j})\bigg[\frac{\cos(\omega_{\bf
{k}}R)}{R^3}+\frac{\omega_{\bf{k}}\sin(\omega_{\bf
{k}}R)}{R^2}\bigg]-(\delta_{ij}-\hat{R_i}\hat{R_j})\frac{\omega^2_{\bf
k}\cos(\omega_{\bf{ k}}R)}{R}\;,
\end{equation}
where the distance of the two atoms is denoted by $R=|\bf
{r_A}-\bf {r_B}|$ and $\hat{R_i}={R_i}/R$ denotes the $i$-th element
of the unit displacement vector of ${\bf{R}}/R\;$.
 In the far zone ($R\omega_{0}\gg1$),  the retardation effect
becomes significant, and we can replace the  dynamical  polarizabilities
$\alpha_{A,B}(\omega_{\bf{k}})$ with  their static polarizabilities
$\alpha_{A,B}(\omega_{\bf{k}})\simeq\alpha_{A,B}(0)\;$~\cite{Spagnolo,Milonni}.
Then  we can write Eq.~(\ref{VAB}) in the far zone as
\begin{equation}\label{VAB-2}
V_{AB}=\alpha_A(0)\alpha_B(0)\int\sum_{ij}\langle {D_i(\omega_{\bf{ k}},{\bf{r_A}})}{D_j(\omega_{\bf
{k}},{\bf {r_B}})}\rangle{V}_{ij}(\omega_{\bf{ k}},R)d\omega_{\bf
{k}}\;.
\end{equation}

 For a Schwarzschild black hole, there are three vacuum states  which can be
defined  by the nonoccupation of positive frequency modes, i.e., the
Boulware, Hartle-Hawking and Unruh vacua.  In the following, we will examine in detail
Eq.~(\ref{VAB-2}) in these vacuum states outside  a Schwarzschild
black hole.

\section{the interatomic Casimir-Polder  potential in Boulware vacuum}
\begin{figure}[htbp]\centering
\includegraphics[scale=0.8]{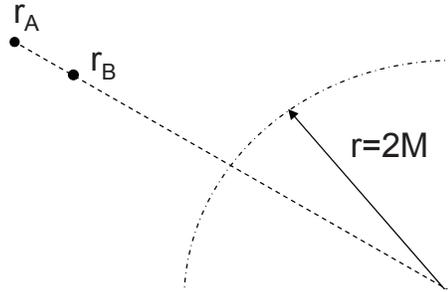}
\caption{The dashed arc denotes the event horizon of a black hole.
Suppose that atom A and atom B are fixed along the same radial
direction, then $R=r_A-r_B\;.$ }\label{tu1}
\end{figure}

Consider the two atoms in interaction with vacuum electromagnetic
fluctuations outside a four-dimensional spherically symmetric black
hole.  The line element of the space-time is given by
\begin{equation}
ds^2=g_{\mu\nu}dx^{\mu}dx^{\nu}=(1-2M/r)dt^2-(1-2M/r)^{-1}dr^2-r^2(d\theta^2+\sin^2\theta{d\phi^2})\;,
\end{equation}
where M is the mass of the black hole.
 Now we suppose that the field is in a vacuum state, and for simplicity, the two
atoms are fixed along the same radial direction(see Fig.~(\ref{tu1}
)).  Then we do not need to
calculate the contributions of spatial field correlation  function
in $\theta-$ and $\phi-$directions.  In this case, Eq.~(\ref{VAB-2}) can be
simplified as
\begin{eqnarray}\label{VAB-2-2}
V_{AB}=\alpha_A(0)\alpha_B(0)\int \langle {D_r(\omega_{\bf{
k}},r_A)}{D_r(\omega_{\bf{ k}},r_B)}\rangle{V}_{rr}(\omega_{\bf
{k}},R)d\omega_{\bf {k}} \;,
\end{eqnarray}
with
\begin{equation}
{V}_{rr}(\omega_{\bf{ k}},R)=-2\bigg[\frac{\cos(\omega_{\bf
{k}}R)}{R^3}+\frac{\omega_{\bf {k}}\sin(\omega_{\bf
{k}}R)}{R^2}\bigg]\;.
\end{equation}

 For the case of the Boulware vacuum,
the  two point  function has been given in Ref.~\cite{ZhouYu-1}
\begin{eqnarray}\label{Boulware concrete two point function}
 \langle{D}_r(x_A){D}_r(x_B)\rangle&=&\frac{1}{4\pi}\sum_{l}
    \int_{0}^\infty d\omega\;\omega\;
    e^{-i\omega(t-t')}(2l+1)\;
    \nonumber\\&&\quad\;\quad\;\;\times
    [\overrightarrow{R}_l(\omega|r_A)\overrightarrow{R}^{\star}_l(\omega|r_B)+
     \overleftarrow{R}_l(\omega|r_A)\overleftarrow{R}^{\star}_l(\omega|r_B)]\;,
\end{eqnarray}
where $\overrightarrow{R}_l$ and $ \overleftarrow{R}_l$ represent
the auxiliary radial function of the outgoing modes from the past
horizon $H^{-}$  and the incoming modes from the past null infinity
$\mathcal{J}^{-}$ respectively~\cite{Luis}. Here, the constant
coefficient is different from that given in Ref.~\cite{ZhouYu-1}
because of the different unit systems. Besides these, it should be
pointed out that  $\omega$ in Eq.~(\ref{Boulware concrete two point
function}) is concerned with the coordinate time $t$. However, for
the atom fixed at a point of a static space-time, the proper
frequency $\omega_{\bf{k}}$ should be associated with the proper
time $\tau$ in the local inertial frame of the atom.
For the case of $r_A,r_B\rightarrow\infty $, it is easy to obtain
that $g_{00}^A\simeq g_{00}^B=g_{00}\sim1\;.$
 When  two atoms are fixed near the event horizon, we will assume
that the distance of the system from the event
horizon is much larger than  the size of the two-atom system itself,
i.e., $R/L\ll1,\;{L}/(2M)\ll1$ with $L=r_B-2M$,  then
$g_{00}^A\simeq g_{00}^B=g_{00}\;$.   Consequently,  the equal-time
(proper time $\tau$) correlation function can be obtained by using
the relation of $\omega=\sqrt{g_{00}}\omega_{\bf{k}}$,  since our
discussions will be focused on  two asymptotic regions, i.e, at the
spatial infinity and near the event horizon. So, we have
\begin{eqnarray}
\langle {D_r(\omega_{\bf{ k}},r_A)}{D_r(\omega_{\bf
{k}},r_B)}\rangle=&&\frac{g_{00}}{4\pi}\sum_{l}{\omega_{\bf{
k}}}(2l+1)\big[
\overrightarrow{R}_l(\omega_{\bf{k}}\sqrt{g_{00}}|r_A)\overrightarrow{R}^{\star}_l(\omega_{\bf
{k}}\sqrt{g_{00}}|r_B)\nonumber\\&&+
     \overleftarrow{R}_l(\omega_{\bf{ k}}\sqrt{g_{00}}|r_A)\overleftarrow{R}^{\star}_l(\omega_{\bf
     {k}}\sqrt{g_{00}}|r_B)]\;.
\end{eqnarray}

Then Eq.~(\ref{VAB-2-2}) can be evaluated by using the corresponding
correlation function
\begin{eqnarray}\label{VAB-3}
V_{AB}=&&\frac{g_{00}\alpha_A(0)\alpha_B(0)}{4\pi}\int_{0}^\infty
\sum_l{\omega_{\bf{ k}}}(2l+1)\big[ \overrightarrow{R}_l(\omega_{\bf
{k}}\sqrt{g_{00}}|r_A)\overrightarrow{R}^{\star}_l(\omega_{\bf
{k}}\sqrt{g_{00}}|r_B)\nonumber\\&&+
     \overleftarrow{R}_l(\omega_{\bf{ k}}\sqrt{g_{00}}|r_A)\overleftarrow{R}^{\star}_l(\omega_{\bf{ k}}\sqrt{g_{00}}|r_B)\big]
     \times{V}_{rr}(\omega_{\bf {k}},R)d\omega_{\bf{ k}}\;.
\end{eqnarray}

It is a formidable task to give the exact forms  of the auxiliary
radial functions. However, the summation concerned with the radial
functions in the two asymptotic regions behaves as (see Appendix)
 \begin{equation}
\sum_l(2l+1)\;\overleftarrow{R}_l(p|r_A)\overleftarrow{R}^{\star}_l(p|r_B)\sim
 \left\{
   \begin{array}{ll}
     \frac{\sum_{l}l(l+1)(2l+1)\;|\mathcal{T}_l(p)|^2}{(2M)^4\;p^2}e^{-ip\Delta{r_*}},
          \;\quad\;\quad\;&r_A,r_B\sim2M\;, \\
     \frac{8\sin(p R/\sqrt{g_{00}})}{\sqrt{g_{00}}R^3p}
     -\frac{8\cos(p R/\sqrt{g_{00}})}{R^2g_{00}}\;,
         \;\;\;\;&r_A,r_B\rightarrow\infty\;,
   \end{array}
 \right.\label{asymptotic summation of ingoing modes}
 \end{equation}
and
 \begin{equation}
\sum_l(2l+1)\;\overrightarrow{R}_l(p|r_A)\overrightarrow{R}^{\star}_l(p|r_B)\sim
 \left\{
   \begin{array}{ll}
     \big(8p^2+\frac{1}{2M^2}\big)\big[\frac{\sin(p{R}/\sqrt{g_{00}})}{\sqrt{g_{00}}p^3R^3}
-\frac{\cos(p{R}/\sqrt{g_{00}})}{g_{00}p^2R^2}\big]\;,
    & r_A,r_B\sim2M\;, \\
     \frac{\sum_{l}l(l+1)(2l+1)\;|\mathcal{T}_l(p)|^2}{{p}^2 r_A^2r_B^2}e^{ip\Delta{r_*}}\;,
      \quad\;\quad\;&r_A,r_B\rightarrow\infty\;,
   \end{array}
 \right.\label{asymptotic summation of outgoing modes}
 \end{equation}
 where $R=r_A-r_B\;$ and $\Delta{r_*}=r_*^A-r_*^B$, with the Regge-Wheeler tortoise
 coordinate defined by $r_*=r+2M\ln(r/2M-1).$
For the sake of  convenience, we  divide the Casimir-Polder
potential into two parts:
$
V_{AB}={\overrightarrow{V}}_{AB}+{\overleftarrow{V}}_{AB}\;,
$
where the contribution of the outgoing modes is denoted by
 \begin{equation}
 {\overrightarrow{V}}_{AB}=\frac{g_{00}\alpha_A(0)\alpha_B(0) }{4\pi}\sum_l\int_{0}^\infty{\omega_{\bf{ k}}}
(2l+1) \overrightarrow{R}_l(\omega_{\bf
{k}}\sqrt{g_{00}}|r_A)\overrightarrow{R}^{\star}_l(\omega_{\bf
{k}}\sqrt{g_{00}}|r_B){V}_{rr}(\omega_{\bf{ k}},R)d\omega_{\bf
{k}}\;,
\end{equation}
and that of the incoming modes by
\begin{equation}
{\overleftarrow{V}}_{AB}=\frac{g_{00}\alpha_A(0)\alpha_B(0)}{4\pi}\sum_l\int_{0}^\infty{\omega_{\bf
{k}}} (2l+1) \overleftarrow{R}_l(\omega_{\bf
{k}}\sqrt{g_{00}}|r_A)\overleftarrow{R}^{\star}_l(\omega_{\bf
{k}}\sqrt{g_{00}}|r_B){V}_{rr}(\omega_{\bf {k}},R)d\omega_{\bf
{k}}\;.
\end{equation}
Using Eq.~(\ref{asymptotic summation of ingoing modes}) and
Eq.~(\ref{asymptotic summation of outgoing modes}), we can show that
${\overleftarrow{V}}_{AB}$ can be approximated at spatial infinity
as
\begin{equation}\label{Boulware-infty-in}
 {\overleftarrow{V}}_{AB}\simeq-\frac{5\alpha_A(0)\alpha_B(0)}{2\pi}\frac{1}{R^7}\;,
\end{equation}
 whereas,
\begin{equation}\label{Boulware-infty-out}
 {\overrightarrow{V}}_{AB}\simeq\frac{\alpha_A(0)\alpha_B(0)}{4\pi}Re\big[\int_{0}^\infty
f(\omega_{\bf {k}},r_A,r_B){V}_{rr}(\omega_{\bf
{k}},R)e^{i\omega_{\bf{k}}{R}\sqrt{
g_{00}}}\omega_{\bf{k}}^3d\omega_{\bf {k}}\big]\;,
\end{equation}
in which
\begin{equation}
 f(\omega_{\bf {k}},r_A,r_B)=\frac{\sum_{l}l(l+1)(2l+1)\;|\mathcal{T}_l(\omega_{\bf{k}}\sqrt{g_{00}})|^2}{\omega_{\bf {k}}^4 r_A^2r_B^2}\
\end{equation}
is a grey-body factor that characterizes the backscattering of the
electromagnetic field modes off the space-time
curvature~\cite{ZhouYu-1}. This grey-body factor is dependent on the
transmission coefficients
$|\mathcal{T}_l(\omega_{\bf{k}}\sqrt{g_{00}})|^2\;$ defined in Ref.~\cite{ZhouYu-1}, of which the exact
analytic expression is not easy to obtain.
However, one can show that by using geometrical optics approximation
and quantum tunneling, the transmission coefficients can be
approximated as \cite{Fabbri,dewitt}
\begin{equation}
|\mathcal{T}_l(\omega_{\bf{k}}\sqrt{g_{00}})|^2\sim
 \left\{
   \begin{array}{ll}
     \theta(\sqrt{27}M\omega_{\bf{k}}\sqrt{g_{00}}-l),\;\quad\;\quad\;&M\omega_{\bf{k}}\gg1\;, \\
   4\big[\frac{(l+1)!(l-1)!}{(2l)!(2l+1)!!}\big]^2(2M\omega_{\bf{k}}\sqrt{g_{00}})^{2l+2}\;,
         \;\;\;\;&M\omega_{\bf{k}}\ll1\;,
   \end{array}
 \right.
 \end{equation}
 where $\theta(x)$ represents the Heaviside $\theta$ function.
Therefore, the grey-body factor may be approximately written as
$f(\omega_{\bf {k}},r_A,r_B)\propto 8g_{00}^2M^4/(3r_A^2r_B^2)\;.$
As a result,
${\overrightarrow{V}}_{AB}\sim0$  at spatial infinity.


However, when the two atoms are fixed near the event horizon ,
 the  leading terms from the contribution of the outgoing modes
 become
\begin{equation}\label{Boulware-2M-out}
{\overrightarrow{V}}_{AB}\simeq
-\frac{5\alpha_A(0)\alpha_B(0)}{2\pi}\frac{1}{R^7}-\frac{3\alpha_A(0)\alpha_B(0)}{16{\pi}M^2g_{00}}\frac{1}{R^5}\;.
\end{equation}
 Let us note here that $M^2g_{00}\gg{R^2}$ since we assume $R/L\ll1,{L}/(2M)\ll1$.  If the size of the two-atom system is not negligible as
 compared with its distance from the event horizon(i.e., $R/L\ll1\;$ is not
 satisfied), then
 we can not take $ g_{00}^B\simeq{g_{00}^A}$.
Physically, this means that  the classical potential tensor of
 the induced dipoles Eq.~(\ref{Vij}) can not be  established because
 of
 the oscillations of the two induced dipoles  at significantly different proper
 frequencies.  We can also show that the contribution from the  incoming modes  behaves as
\begin{equation}\label{Boulware-2M-in}
{\overleftarrow{V}}_{AB}\simeq
\frac{\alpha_A(0)\alpha_B(0)}{4\pi}Re\big[\int_{0}^\infty
{f(\omega_{\bf {k}},r_A,r_B)} {V}_{rr}(\omega_{\bf
{k}},R)e^{-i\omega_{\bf{k}}R\sqrt{g_{00}}}\omega_{\bf{k}}^3d\omega_{\bf
{k}}\big]\;,
\end{equation}
which is much smaller than  Eq.~(\ref{Boulware-2M-out})
 as a result of the vanishingly-small grey-body factor
 near the even horizon. In summary, the interatomic
Casimir-Polder potential in the Boulware vacuum is given by
 \begin{equation}\label{Boulware-Vab}
V_{AB}\simeq
  \left\{
     \begin{array}{ll}
   -\frac{5\alpha_A(0)\alpha_B(0)}{2\pi}\frac{1}{R^7}\;,
   \quad\;\quad\;&r_A, r_B\rightarrow\infty\;, \\
-\frac{5\alpha_A(0)\alpha_B(0)}{2\pi}\frac{1}{R^7}
-\frac{3\alpha_A(0)\alpha_B(0)}{16{\pi}M^2g_{00}}\frac{1}{R^5}\;,
  \quad\;\quad\; &r_A,r_B\sim2M\;.
 \end{array}
      \right.
 \end{equation}

\section{the interatomic Casimir-Polder  potential in Hartle-Hawking vacuum}
 For the case of the Hartle-Hawking vacuum, Eq.~(\ref{VAB-2-2})
can  also  be written as
\begin{eqnarray}\label{VAB-4}
V_{AB}&=&\frac{g_{00}\alpha_A(0)\alpha_B(0)}{4\pi}\int_{-\infty}^\infty\sum_l{\omega_{\bf{
k}}} (2l+1)\bigg[ \frac{\overrightarrow{R}_l(\omega_{\bf
{k}}\sqrt{g_{00}}|r_A)\overrightarrow{R}^{\star}_l(\omega_{\bf
{k}}\sqrt{g_{00}}|r_B)}{1-e^{-\omega_{\bf {k}}/T}}\nonumber\\&&+
    \frac{ \overleftarrow{R}^{\star}_l(\omega_{\bf {k}}\sqrt{g_{00}}|r_A)
    \overleftarrow{R}_l(\omega_{\bf {k}}\sqrt{g_{00}}|r_B)}{e^{\omega_{\bf {k}}/T}-1}\bigg]
     \times{V}_{rr}(\omega_{\bf {k}},R)d\omega_{\bf {k}}\;,
\end{eqnarray}
where $T=T_H/\sqrt{g_{00}}$ with $T_H=1/(8\pi{M})$ being the usual
Hawking temperature ~\cite{ZhouYu-1}. With  the help of the
approximate forms of the radial functions in the two asymptotic
regions, Eq.~(\ref{VAB-4}) can be evaluated, in the case of $r_A,
{r_B}\rightarrow\infty$ , to get
\begin{equation}\label{Hartle-infty-in}
{\overleftarrow{V}}_{AB}\simeq-\frac{4\pi^2{T^3}\alpha_A(0)\alpha_B(0)\coth(2\pi{R}T)}{R^4\sinh^2(2\pi{R}T)}
-\frac{4{\pi}T^2\alpha_A(0)\alpha_B(0)}{{R^5}\sinh^2(2{R}T)}
-\frac{2T\alpha_A(0)\alpha_B(0)\coth(2\pi{R}T)}{{R^6}}\;,
\end{equation}
and
\begin{equation}\label{Hartle-infty-out}
{\overrightarrow{V}}_{AB}\simeq\frac{\alpha_A(0)\alpha_B(0)}{4\pi}Re\big[\int_{-\infty}^\infty
\frac{f(\omega_{\bf {k}},r_A,r_B)}{1-e^{-\omega_{\bf
{k}}/T}}{V}_{rr}(\omega_{\bf
{k}},R)e^{i\omega_{\bf{k}}R\sqrt{g_{00}}}\omega_{\bf{k}}^3d\omega_{\bf
{k}}\big]\;.
\end{equation}

At spatial infinity ($r_A, {r_B}\rightarrow\infty$),
${\overleftarrow{V}}_{AB}$ is the dominant term compared
with${\overrightarrow{V}}_{AB}\sim{r_A^{-2}r_B^{-2}}\;.$
Therefore, the interatomic Casimir-Polder potential can be
simplified further by only considering ${\overleftarrow{V}}_{AB}$
\begin{equation}\label{Hartle-infty-Vab}
V_{AB}\simeq
  \left\{
     \begin{array}{ll}
   -\frac{5\alpha_A(0)\alpha_B(0)}{2\pi}\frac{1}{R^7}-\frac{8\pi^3\alpha_A(0)\alpha_B(0)}{45}\frac{T_H^4}{R^3}\;,
   \quad\;\quad\;& T_HR\ll1\;, \\
   -\frac{2T_H\alpha_A(0)\alpha_B(0)}{R^6},
  \quad\;\quad\; & T_HR\gg1\;,
 \end{array}
      \right.
 \end{equation}
 where $T\sim{T_H}$ is taken at spatial infinity.
 It is obvious  to see that  $V_{AB}$  is similar to the Casimir-Polder potential at finite temperature ~\cite{Goedecke,Spagnolo-2,Boyer-2}.
 This result is consistent with our usual understanding that the Hartle-Hawking vacuum describes
a black hole in equilibrium with an infinite sea of black-body
radiation at Hawking temperature. When comparing
Eq.~(\ref{Hartle-infty-out}) with Eq.~(\ref{Boulware-infty-out}), we
find out that Eq.~(\ref{Hartle-infty-out}) is dependant on the
temperature $T\;$. This is in accordance with the common belief that
thermal flux emanates from the black hole  which is partly depleted
by backscattering off the space-time  curvature on its way to infinity.

In the region near the event horizon of a black hole (i.e.,
$R/L\ll1,{L}/(2M)\ll1\;$), the contribution from the incoming modes
behaves as
\begin{equation}\label{Hartle-2M-in}
 {\overleftarrow{V}}_{AB}\simeq\frac{\alpha_A(0)\alpha_B(0)}{4\pi}Re\big[\int_{-\infty}^\infty
\frac{f(\omega_{\bf {k}},r_A,r_B)}{e^{\omega_{\bf
{k}}/T}-1}{V}_{rr}(\omega_{\bf
{k}},R)e^{i\omega_{\bf{k}}R\sqrt{g_{00}}}\omega_{\bf{k}}^3d\omega_{\bf
{k}}\big]\;.
 \end{equation}
Obviously, ${\overleftarrow{V}}_{AB}$ is vanishingly small  due to
the grey-body factor. Then the interatomic Casimir-Polder potential
 is mainly determined by ${\overrightarrow{V}}_{AB}\;.$
When $R/L\ll1,{L}/(2M)\ll1\;$, it is easy to
deduce that $T R\ll1\;.$  Then we find
\begin{equation}\label{Hartle-2M}
V_{AB}\simeq -\frac{5\alpha_A(0)\alpha_B(0)}{2\pi}\frac{1}{R^7}
-\frac{3\alpha_A(0)\alpha_B(0)}{16{\pi}M^2g_{00}}\frac{1}{R^5}-\frac{\pi\alpha_A(0)\alpha_B(0)}{36M^2g_{00}}\frac{T^2}{R^3}
-\frac{8\pi^3\alpha_A(0)\alpha_B(0)}{45}\frac{T^4}{R^3}\;.
 \end{equation}

According to Eq.~(\ref{Hartle-2M}), it is easy to see that both the
curvature of space-time and the thermal radiation  make the interatomic Casimir-Polder potential smaller.
One can also see that the first  two terms  Eq.~(\ref{Hartle-2M}) are
just the interatomic Casimir-Polder potential near the horizon in the Boulware vacuum (Eq.~(\ref{Boulware-Vab})) and the
last two terms can be considered as the contribution of the Hawking radiation of the black hole.

\section{the interatomic Casimir-Polder  potential in Unruh vacuum}

For the case of Unruh vacuum, the far-zone interatomic
Casimir-Polder potential of two ground-state atoms becomes~\cite{ZhouYu-1}
\begin{eqnarray}\label{VAB-5}
V_{AB}&=&\frac{g_{00}\alpha_A(0)\alpha_B(0)}{4\pi}\int_{-\infty}^\infty
\sum_l{\omega_{\bf {k}}}(2l+1)\big[
\frac{\overrightarrow{R}_l(\omega_{\bf
{k}}\sqrt{g_{00}}|r_A)\overrightarrow{R}^{\star}_l(\omega_{\bf
{k}}\sqrt{g_{00}}|r_B)}{1-e^{-\omega_{\bf {k}}/T}}\nonumber\\&&+
  \theta(\omega_{\bf {k}}){ \overleftarrow{R}_l(\omega_{\bf {k}}\sqrt{g_{00}}|r_A)\overleftarrow{R}^{\star}_l(\omega_{\bf {k}}\sqrt{g_{00}}|r_B)}\big]
     \times{V}_{rr}(\omega_{\bf {k}},R)d\omega_{\bf {k}}\;.
\end{eqnarray}

Similarly, we can also obtain the
approximate results in the two asymptotic regions. When   two atoms
are fixed at spatial infinity, the contribution of  the outgoing modes is the
same as Eq.~(\ref{Hartle-infty-out}), which is negligible, and then the corresponding Casimir-Polder
interatomic potential is mainly determined by  the contribution from the incoming modes
(similar to Eq.~(\ref{Boulware-infty-in}))
\begin{eqnarray}
V_{AB}\simeq
    -\frac{5\alpha_A(0)\alpha_B(0)}{2\pi}\frac{1}{R^7}\;.
\end{eqnarray}

When two atoms are fixed near the horizon, the contribution from the incoming modes   which reads
 \begin{equation}
 {\overleftarrow{V}}_{AB}\simeq\frac{\alpha_A(0)\alpha_B(0)}{4\pi}Re[\int_{0}^\infty
f(\omega_{\bf {k}},r_A,r_B){V}_{rr}(\omega_{\bf
{k}},R)e^{-i\omega_{\bf{k}}R\sqrt{g_{00}}}\omega_{\bf{k}}^3d\omega_{\bf
{k}}\big]\;,
 \end{equation}
 is the vanishingly small  and  the dominant term of the interatomic Casimir-Polder potential arises
from the contribution of the outgoing modes. This situation is similar to the case of the
Hartle-Hawking vacuum. We then have
\begin{equation}\label{Unruh-2M}
V_{AB}\simeq  -\frac{5\alpha_A(0)\alpha_B(0)}{2\pi}\frac{1}{R^7}
-\frac{3\alpha_A(0)\alpha_B(0)}{16{\pi}M^2g_{00}}\frac{1}{R^5}-\frac{\pi\alpha_A(0)\alpha_B(0)}{36M^2g_{00}}\frac{T^2}{R^3}
-\frac{8\pi^3\alpha_A(0)\alpha_B(0)}{45}\frac{T^4}{R^3}\;.
 \end{equation}

Therefore, we conclude that at spatial infinity the interatomic
Casimir-Polder potential in the Unruh vacuum is the same as that in the Boulware vacuum with a $R^{-7}$ behavior and the contribution
of the outgoing thermal radiation is negligible as a result of the backscattering off the spacetime on its way to infinity. When the two atoms are fixed near the horizon,
the corresponding  far-zone interatomic Casimir-Polder potential is
the same as  that in the Hartle-Hawking vacuum.

\section{Conclusion}
In this paper, we have studied the far-zone interatomic
Casimir-Polder potential between two atoms outside a Schwarzschild
black hole. We find that at spatial infinity, the behavior of the
Casimir-Polder potential in the Boulware vacuum is similar to that
in  vacuum in a flat spacetime with a $R^{-7}$ behavior, and the same is true for the Casimir-Polder potential in
the Unruh vacuum  as a result of the backscattering of the Hawking
radiation from the black hole off the spacetime curvature. However,
the Casimir-Polder potential in Hartle-Hawking vacuum behaves like
that in a  thermal bath at the Hawking temperature. Close to the
event horizon, the space-time curvature induces modifications to the
interatomic Casimir-Polder potential in all three vacuum states,
making the potential smaller.

\newpage

\appendix*
\section{the summation concerning the radial functions}
In order to prove  Eq.~(\ref{asymptotic summation of ingoing modes})
and Eq.~(\ref{asymptotic summation of outgoing modes}), we first
introduce some  conclusions  in  Ref.~\cite{ZhouYu-1}. In the Boulware
vacuum, the two point correlation function of electromagnetic fields
reads
\begin{eqnarray}\label{two point function}
\langle{D}_r(x_A){D}_r(x_B)\rangle&=&\frac{g_{00}}{4\pi}
    \int_{0}^\infty d\omega_{\bf{k}}\;\omega_{\bf{k}}\;
    e^{-i\omega_{\bf{k}}\Delta\tau}\sum_{l}(2l+1)[\overrightarrow{R}_l(\omega_{\bf{k}}\sqrt{g_{00}}|r_A)
    \overrightarrow{R}^{\star}_l(\omega_{\bf{k}}\sqrt{g_{00}}|r_B)
  \nonumber\\&&+
  \overleftarrow{R}_l(\omega_{\bf{k}}\sqrt{g_{00}}|r_A)\overleftarrow{R}^{\star}_l(\omega_{\bf{k}}\sqrt{g_{00}}|r_B)]\;,
\end{eqnarray}
with
 \begin{equation}
R^{(n)}_{l}(\omega
|r)=\frac{\sqrt{l(l+1)}}{\omega}\frac{\varphi^{(n)}_{\omega
l}(r)}{r^2}\;.
\end{equation}
Here the label $``n"$ distinguishes between incoming  modes (denoted
with $n=\leftarrow$) and outgoing modes(denoted with
$n=\rightarrow$). The asymptotic expressions of the radial function
in the two asymptotic regions single out
 \begin{eqnarray}
&&\overrightarrow{\varphi}_{\omega l}(r)\sim
  \left\{
    \begin{array}{ll}
      e^{i\omega r_*}+\overrightarrow{\mathcal{R}_l}(\omega)\;e^{-i\omega r_*}\;,
             \quad\;\quad\;r\sim 2M\;, \\
      \overrightarrow{\mathcal{T}_l}(\omega)\;e^{i\omega r_*},
             \quad\;\quad\;\quad\;\quad\;\quad\;\;r\rightarrow\infty\;,
    \end{array}
  \right.\label{asymptotic outgoing mode}\\
&&\overleftarrow{\varphi}_{\omega l}(r)\sim
 \left\{
    \begin{array}{ll}
      \overleftarrow{\mathcal{T}_l}(\omega)\;e^{-i\omega r_*}\;,
             \quad\;\quad\;\quad\;\quad\quad r\sim2M\;, \\
      e^{-i\omega r_*}+\overleftarrow{\mathcal{R}_l}(\omega)\;e^{i\omega r_*}\;,
             \quad\;\quad\;r\rightarrow\infty\;.
    \end{array}
  \right.\label{asymptotic ingoing mode}
\end{eqnarray}
Here $\mathcal{R}$ and $\mathcal{T}$ are, respectively, the
reflection and transmission coefficients. If $r_A=r_B$, it has been
proven that (see Appendix in Ref.~\cite{ZhouYu-1})
 \begin{equation}
\sum_l(2l+1)\;|\overleftarrow{R}_l(p|r_B)|^2\sim
 \left\{
   \begin{array}{ll}
     \frac{\sum_{l}l(l+1)(2l+1)\;|\mathcal{T}_l(p)|^2}{(2M)^4\;p^2},
          \;\quad\;\quad\;r_A=r_B\sim2M\;, \\
     \frac{8p^2}{3g_{00}^2}\;,\quad\;\quad\;\quad\;\quad\;\quad\;\quad\;\;
         \;\;\;\;r_A=r_B\rightarrow\infty\;,
   \end{array}
 \right.\label{ingoing modes}
 \end{equation}
and
 \begin{equation}
\sum_l(2l+1)\;|\overrightarrow{R}_l(p|r_B)|^2\sim
 \left\{
   \begin{array}{ll}
     \frac{8p^2}{3g_{00}^2}+\frac{1}{6M^2g_{00}^2}\;,
      \quad\;\quad\;\quad\quad\;r_A=r_B\sim2M\;, \\
     \frac{\sum_{l}l(l+1)(2l+1)\;|\mathcal{T}_l(p)|^2}{p^2r_B^4}\;,
      \quad\;\quad\;r_A=r_B\rightarrow\infty\;.
   \end{array}
 \right.\label{outgoing modes}
\end{equation}

At spatial infinity ($r_A,{r_B}\rightarrow\infty$), the equal-time
correlation function Eq.~(\ref{two point function}) should be
identified with the equal-time correlation function in Minkowski
space~\cite{Power-Thirunamachandran}
\begin{eqnarray}\label{d-minkowski}
\langle 0|{D}_i(x_A){D}_j(x_B)|0\rangle=\frac{1}{\pi}
    \int_{0}^\infty d\omega_{\bf{k}}\omega_{\bf{k}}^3e^{-i\omega_{\bf{k}}\Delta\tau}
    \bigg\{(\delta_{ij}-\hat{R_i}\hat{R_j})\frac{\sin(\omega_{\bf{k}}R)}{\omega_{\bf{k}}R}\nonumber\\
    +(\delta_{ij}-3\hat{R_i}\hat{R_j})\bigg[\frac{\cos(\omega_{\bf{k}}R)}{\omega_{\bf{k}}^2R^2}
    -\frac{\sin(\omega_{\bf{k}}R)}{\omega_{\bf{k}}^3R^3}\bigg]\bigg\}\;.
\end{eqnarray}
When comparing Eq.~(\ref{two point function}) with
Eq.~(\ref{d-minkowski}), we  obtain that
\begin{equation}\label{leftomega-k}
 \sum_l(2l+1)\overleftarrow{R}_l(\omega_{\bf{k}}\sqrt{g_{00}}|r_A)\overleftarrow{R}^{\star}_l(\omega_{\bf{k}}\sqrt{g_{00}}|r_B)\simeq
 \frac{8\omega_{\bf{k}}^2}{g_{00}}\bigg[\frac{\sin(\omega_{\bf{k}}R)}{\omega_{\bf{k}}^3R^3}-\frac{\cos(\omega_{\bf{k}}R)}{\omega_{\bf{k}}^2R^2}
    \bigg]\;,
\end{equation}
where the  term about $\overrightarrow{R}_l(\omega_{\bf{k}})$ in
Eq.~(\ref{two point function}) is  neglected because this is very
small at the asymptotic region  due to  outgoing modes backscattered
off the space-time curvature on their way. Through a simple
calculations, we can write Eq.~(\ref{leftomega-k}) as
\begin{eqnarray}\label{leftomega-k2}
\sum_l(2l+1)\overleftarrow{R}_l(p|r_A)\overleftarrow{R}^{\star}_l(p|r_B)&\simeq&\frac{8p^2}{3g_{00}^2}
\bigg[\frac{3\sin(p{R}/\sqrt{g_{00}})}{p^3R^3/\sqrt{g_{00}^3}}
-\frac{3\cos(p{R}/\sqrt{g_{00}})}{p^2R^2/g_{00}}\bigg]\;.
\end{eqnarray}
This agrees with approximative summation relations
Eq.~(\ref{asymptotic summation of ingoing modes}) in the case of
$r_A,r_B\rightarrow\infty$. For the case $r_A,r_B\sim2M$, the
corresponding result is easy to obtain by using
Eq.~(\ref{asymptotic ingoing mode}).

When comparing the expression of $\overrightarrow{\varphi}_{\omega
l}(r)$ near the horizon (Eq.~(\ref{asymptotic outgoing mode})) with
the expression of $\overleftarrow{\varphi}_{\omega l}(r)$ at spatial
infinity(Eq.~(\ref{asymptotic ingoing mode})), we find that
there are some similarities and symmetries among these equations.
According to the relations of Eq.~(\ref{ingoing modes}) (the case of
$r_A=r_B\rightarrow\infty$) and Eq.~(\ref{leftomega-k2}), it is not
difficult to deduce that $r_A,{r_B}\sim{2M}$, satisfying
\begin{eqnarray}\label{rightomega-k}
\sum_l(2l+1)\overrightarrow{R}_l(p|r_A)\overrightarrow{R}^{\star}_l(p|r_B)&\simeq&
\bigg(\sum_l(2l+1)|\overrightarrow{R}_l(p|r_B)|^2\bigg)\bigg[\frac{3\sin(p{R}/\sqrt{g_{00}})}{p^3R^3/\sqrt{g_{00}^3}}
\nonumber\\&&-\frac{3\cos(p{R}/\sqrt{g_{00}})}{p^2R^2/g_{00}}\bigg]\;.
\end{eqnarray}
Therefore, we can prove Eq.~(\ref{asymptotic summation of outgoing
modes}) with  some
 simple calculations.
\begin{acknowledgments}
 This work was supported in part by the National Natural Science
Foundation of China under Grants  No. 11075083, No.11005038, No.
10935013 and No. 11375092; the Zhejiang Provincial Natural Science Foundation of
China under Grant No. Z6100077;  the National Basic Research Program
of China under Grant No. 2010CB832803; the PCSIRT under Grant No.
IRT0964, and the Hunan Provincial Natural Science Foundation of
China under Grant No. 11JJ7001.
\end{acknowledgments}

\end{document}